\documentstyle[12pt,fleqn,epsfig]{article} 

\newcommand{\be}{\begin{equation}} 
\newcommand{\ee}{\end{equation}} 
\newcommand{\bear}{\be\begin{array}} 
\newcommand{\eear}{\end{array}\ee} 
\newcommand{\bea}{\begin{eqnarray}} 
\newcommand{\eea}{\end{eqnarray}} 

\catcode`@=11 
\newtoks\@stequation 
\def\subequations{\refstepcounter{equation}% 
\edef\@savedequation{\the\c@equation}% 
 \@stequation=\expandafter{\theequation}%   %only want \theequation 
 \edef\@savedtheequation{\the\@stequation}% % expanded once 
 \edef\theequation{\theequation}% 
 \setcounter{equation}{0}% 
 \def\theequation{\theequation\alph{equation}}} 
\def\endsubequations{\setcounter{equation}{\@savedequation}% 
  \@stequation=\expandafter{\@savedtheequation}% 
  \edef\theequation{\the\@stequation}\global\@ignoretrue 
\noindent} 
\catcode`@=12

\begin{document} 

\begin{flushright} 
UCLA/00/TEP/16\\   
May 2000 
\end{flushright}
 
\begin{center} 
\bigskip\bigskip  
 
{ \large{\bf Super-Kamiokande 0.07 eV Neutrinos in Cosmology: Hot Dark Matter
and the Highest Energy Cosmic Rays}}\footnote{Talk given at the 
``4th International Symposium on Sources and Detection of Dark Matter in the 
Universe", February 23-25, 2000, Marina del Rey, CA (to appear in its 
proceedings) and at  the  ``Cosmic Genesis and Fundamental Physics" workshop, 
October 28-30, 1999, Sonoma State University, Santa Rosa, CA.}
 
\bigskip\bigskip 
{ Graciela Gelmini$^1$} 

\bigskip\bigskip 
 
${}^{1}${\em Dept. of Physics and Astronomy, UCLA,  Los Angeles, CA 90095-1547}

\end{center}

\begin{abstract}
Relic neutrinos with mass  in the range indicated by Super-Kamiokande results if 
neutrino masses are hierarchial (about 0.07 eV)  are many times deemed too light 
to be cosmologically relevant.
Here we remark that these neutrinos may significantly contribute to the
dark matter of the Universe (with a large lepton asymmetry $L$) and that
their existence might be revealed by the spectrum of ultra high energy
cosmic rays (maybe even in the absence of a large $L$).
\end{abstract}

Super-Kamiokande has provided a strong evidence for the oscillation in 
atmospheric showers of  two neutrino species with masses $m_1$,
$m_2$ and  $\delta m^2 $ = $m_1^2-m_2^2$ = $(1-8) \times 10^{-3}$ eV consisting
mostly of about equal amounts of $\nu_\mu$ and another flavor eigenstate 
neutrino, 
$\nu_\tau$ or a sterile neutrino\cite{SK}.  If
neutrino masses are hierarchial, as those of the other leptons and quarks,
then the heavier of the two oscillating neutrinos, call it $\nu_{\rm SK}$, has a 
mass $m_{\rm SK} =\sqrt{\delta m^2} \simeq 0.07$ eV.

  The possibility of $m_1$ and $m_2$ being much larger than $m_{\rm SK}$ has also
been invoked, in part with the motivation to have $m_1\sim m_2$ of the order of 
eV, in
the  range previously considered  necessary for relic neutrinos to constitute a
cosmologically relevant component of the dark matter in the Universe.  In
fact, with no lepton asymmetry, i.e. with $L_\nu \equiv 
[(n_\nu-n_{\bar\nu})/n_\gamma]=0$, the number density of relic neutrinos and
antineutrinos of each species is $n_\nu=n_{\bar\nu}=3n_\gamma/22 = 56 cm^{-3}$.
With this number density, the contribution of relic neutrinos to the energy
density of the Universe (in units of the critical density) is $\Omega_\nu h^2=
(\sum_i m_{\nu_i}/92$ eV) (here $h\simeq 0.7$ is the Hubble constant in units of
100 km/Mpc sec). This amounts to only to $\Omega_\nu h^2= 0.8 \times 10^{-3}$,
while a value about 10 times larger was considered necessary in the context of 
Cold-Hot Dark Matter (CHDM) models \cite{Gawiser}.

 In Ref.\cite{GK1} A. Kusenko and I  pointed out that if  the lepton 
asymmetry of $\nu_{\rm SK}$ in the Universe is of order one
the neutrinos with $m_{\rm SK}$ can make a significant contribution to the
energy content of the Universe \footnote{This was also pointed out in 
Ref.\cite{PK}. However, in this ref. the neutrino decoupling temperature $T_d$, 
which  increases with increasing values of  $|L_\nu|$ \cite{FKT}\cite{KS} due to 
the effect of Fermi blocking factors, was instead taken to decrease with 
$|L_\nu|$,  which lead to an incorrect relation between lepton number and 
density.}
\begin{equation}
\Omega_\nu h^2 \simeq 0.01 \left ( \frac{|L_\nu| }{3.6} \right )
\left ( \frac{m_\nu}{0.07 eV} \right ), 
\label{etaomega}
\end{equation}
In the same reference \cite{GK1} A. Kusenko and I also pointed out that, not only 
Fermi-degenerate relic neutrinos with $m_{_{SK}}$ could  be revealed in the 
highest energy cosmic rays (if these are due to ``Z-bursts") but, based on the 
results of Adams and Sarkar\cite{AS}, these neutrinos could be a new form of Hot 
Dark Matter. Adams and Sarkar had found that a massless relic neutrino species 
with chemical potential $\mu_\nu = 3.4~T_\nu$ added to a `standard' Cold Dark 
Matter  (CDM) model (flat matter dominated Universe with no cosmological 
constant) provides a good fit to the large scale structure (LSS) and  anisotropy 
of the cosmic microwave background radiation (CMBR) data.  

 Neutrinos with mass $m_{\rm SK}$, contrary to the neutrinos studied by Adams and 
Sarkar, are non-relativistic at present, however they are still relativistic at 
the time of radiation-matter equality (the time at which the Universe becomes 
matter dominated), when their effect is most important.  

 In fact, the results of Adams and Sarkar were later confirmed and expanded by 
Lesgourgues and Pastor\cite{LP1} who studied the impact of both massless and 
massive relic neutrinos with  $m_{\rm SK}$,  with large lepton asymmetries  on 
structure formation and the CMBR anisotropy. They studied  CDM
 models with and without cosmological constant $\Lambda$.  I propose to call 
these models {\bf LCDM} and {\bf L${\bf \Lambda}$CDM}, where L stands for the 
addition of a large lepton asymmetry.   The major effect of the lighter and more 
abundant relic neutrinos is to delay the onset of matter domination
in the Universe (which increases the amplitudes of the acoustic peaks in the 
angular spectrum of CMBR anisotropies).  The neutrino mass is thus largely 
irrelevant, while neutrinos are relativistic at the moment of radiation-matter 
equality.

For large $L_\nu$ the relation between this asymmetry and the chemical
potential $\xi \equiv \mu_\nu/T_\nu$ (which is constant after neutrinos decouple) is
\begin{equation}
L_\nu= \frac{1}{12 \zeta(3)} \left ( \frac{T_\nu}{T_\gamma}\right )^3
[\pi^2 \xi + \xi^3] = 0.0252 ~(9.87 \xi+\xi^3).
\label{etaxi}
\end{equation}
Here, $\zeta(3)=1.202$ and $(T_\nu/T_\gamma)^3 = 4/11$. This
value for the temperature ratio is valid as long as the neutrino decoupling 
temperature $T_d$ is lower than the muon mass, which translates into the upper 
bound $\xi < 12$~\cite{KS}.

 Neutrinos with chemical potentials $\xi \geq 1$ ($L_\nu \geq 0.27)$
are Fermi-degenerate.  Only for these values of $\xi$ the number density
of relic neutrinos becomes considerably different than in the usual case
with no asymmetry.

 After neutrino-antineutrino annihilation ceases in the early Universe, only the 
particles in excess  remain and  $|L_\nu|$ is just the ratio of the number 
density of these particles,  say $n_\nu$, over the photon density, $|L_\nu |= 
n_\nu/n_\gamma$.  For $\xi = 5$, for example, one obtains $L_\nu = 4$, which 
means 
that there are 4 background neutrinos for every background photon (thus, 
neutrinos dominate the entropy of the Universe) and consequently, $n_\nu \simeq 
1700~{\rm cm}^{-3}$.  This would make relic neutrinos 30 times more abundant
than standard neutrinos or antineutrinos of every species with no lepton
asymmetry and would make the relic density $\Omega_\nu h^2 \simeq (m_\nu/3$ eV).

 Lesgourgues and Pastor\cite{LP1} found that, even with no cosmological constant 
(an assumption disfavored at present by type IA  Supernovae data), i.e. in a {\bf 
LCDM} model, $\nu_{\rm SK}$ with chemical potentials 
$\xi$ between 3 and 6 ($L_\nu$ between 1.4 and 6.9) added to CDM provide a good 
agreement with LSS and CMBR observations. In {\bf L${\bf \Lambda}$CDM} models instead, with a 
cosmological constant contribution of $\Omega_\Lambda = 0.5$ to the energy 
density, $\xi$ could be between 0 to 4 ($L_\nu$ between 0 and 2.6). As the 
cosmological constant increases there is less room for neutrinos and  for 
$\Omega_\Lambda > 0.7$  no-lepton asymmetry is allowed, i.e. $\xi = 0$.  
Kinney and Riotto\cite{KR} also studied
the CMB anisotropy in the presence of large lepton asymmetries, with similar
results.

The large lepton asymmetries invoked here may seem odd.  However, they have
been studied several times from 1967 \cite{Wagoner} onwards, in the context of 
nucleosynthesis.
In the presence of neutrino degeneracy nucleosynthesis becomes severely 
non-standard. A large number of electron neutrinos, $\nu_e$, present during 
nucleosynthesis yields a reduction of the neutron to proton ratio, $n/p$, through 
the reaction $n~\nu_e\to p~e$. This in turn lowers the $^4$He abundance, since 
when nucleosynthesis takes place essentially 
all neutrons end up in $^4$He nuclei.  Extra neutrinos of any flavor increase
the energy density of the Universe, leading to an earlier decoupling of weak
interactions and consequent increase of the $n/p$ ratio
(and $^4$He).  This last effect is less important than the former one in the
case of $\nu_e$, but it is the only effect of an excess of $\nu_\mu$ and/or
$\nu_\tau$ (or their antinuetrinos).  Thus, when both the chemical potentials of 
$\nu_e$ and of $\nu_\mu$ or $\nu_\tau$ are large, their effects largely 
compensate each other. 

 Combining nucleosynthesis bounds with the requirement of neutrinos becoming
subdominant before the recombination epoch, in 1992 Kang and Steigman\cite{KS} 
found 
\begin{equation}
-0.06 \leq \xi_{\nu_e} \leq 1.1,~~|\xi_{{\nu_\mu},{\nu_\tau}}|\leq 
6.9,~~~\Omega_Bh^2\leq 0.069.
\label{Nuc}
\end{equation}
The last is a bound on the baryon density $\Omega_B$
almost an order of magnitude larger than obtained in conventional
nucleosynthesis. Newer data on primordial element abundances  seem to impose 
similar bounds on neutrino chemical potentials \cite{AF}. Notice that it is the 
upper 
bound on $\xi_{\nu_\mu}$ which is relevant for $\nu_{SK}$.

 The existence of ultra high energy cosmic rays (UHECR) with energies above the 
Greisen-Zatsepin-Kuzmin (GZK) cutoff of about $5\times 10^{19}$ eV, presents a 
problem.  Photons and nucleons with those energies have short attenuation lengths 
and could only come from distances of 100 Mpc or less, while possible 
astrophysical sources for those energetic particles are much farther away.  An 
elegant and economical solution  to this problem, proposed by T. Weiler\cite{W} 
consists of the production of the necessary photons and nucleons close to Earth, 
in the annihilation at the $Z$-resonance of  ultra-high energy neutrinos, 
$\nu_{\rm UHE}$, coming from remote sources, and relic background neutrinos.  
These events were named ``$Z$-bursts"  by T. Weiler.
The $Z$-resonance occurs when the energy of the incoming $\nu_{\rm UHE}$ 
neutrinos is $E_{\nu_{\rm UHE}}= E_{Res}$, 
\begin{equation}
E_{Res} = \frac{M_Z^2}{2~m_\nu} ,
\label{Eres}
\end{equation}
where $m_\nu$ is the mass of the relic neutrinos.  Since galaxy formation 
arguments show  $m_\nu <$ few eV, then $E_{\nu_{\rm UHE}} > 10^{21}$ eV, 
precisely above the GZK cutoff, as needed.  As we see in the equation, in this 
mechanism the energy cutoff $E_{Res}$ of the  UHECR is related to the mass of the 
relic neutrinos,  and this cutoff should be $E_{Res} \simeq 0.6~10^{23}$ eV for
$m_\nu = m_{\rm SK}$. 

Depending on the assumed spectrum of $\nu_{\rm UHE}$
 used, upper bounds on the intensity of the $\nu_{\rm UHE}$
flux can be obtained. These bounds determine the need for an enhancement in the 
density of relic neutrinos above the standard
density of 56 cm$^{-3}$,  to account for the observed flux of UHECR.
In the case of eV mass neutrinos the enhancement could come from gravitational
clustering.  Neutrinos with $m_{\rm SK}$ are too light to cluster significantly,
but, if needed, the density enhancement could come from a large lepton asymmetry.

Most bounds on  ``$Z$-burst" models (see for example \cite{bounds}) assume that 
the $\nu_{\rm UHE}$ have a typical astrophysical spectrum, decreasing with
energy as $E^{-\gamma}$, with $\gamma$ a number of order one.  These bounds would
not hold if the $\nu_{\rm UHE}$ spectrum had a very different energy dependence, 
as, for example if the sources would be unstable superheavy relic particles, 
which form part of the cold dark matter, decaying
mostly into neutrinos\cite{GK2}, \cite{sigl}. In this case the spectrum of 
$\nu_{\rm UHE}$  is opposite to an astrophysical spectrum,
it grows rapidly with energy, up to a sharp cutoff at an energy of the order
of the parent particle mass.  A model for these parent particles is arguably 
difficult to obtain\cite{CDF}, but the consequences of this idea make it worth
considering.  
Besides producing, as already mentioned, a spectrum of
$\nu_{\rm UHE}$ very different than those of astrophysical origin, with almost no 
neutrinos at low energy where bounds exist, this idea implies that the directions 
of UHECR could map the distribution of parent particles (which should coincide 
with the distribution of cold dark matter) at large red shifts.  This is because 
the initial energy of the $\nu_{\rm UHE}$ decay product needs to be
redshifted to the energy of the ``$Z$-burst" in its way to the Earth.  This
idea of unstable superheavy relic particles could be constrained by the
EGRET bound on the diffuse low-energy gamma ray flux resulting from the 
$Z$-bursts \cite{S1}.  This question deserves further study. 

In an upcoming
paper\cite{GK3} simulations will be presented for the photon, nucleon and 
neutrino 
fluxes coming from ``$Z$-bursts" of $0.6 \times 10^{23}$ eV, as would be if due to relic 
neutrinos of mass $m_{\rm SK}$.  The ``$Z$-bursts" were simulated using PYTHIA 
and the absorption of photons and nucleons was modelled using energy attenuation
lengths provided by Bhattacharjee and Sigl\cite{BS} supplemented by 
radio-background models by Protheroe and Biermann\cite{PB} . Fig. 1
\begin{figure}
\epsfxsize=\textwidth 
\epsfbox{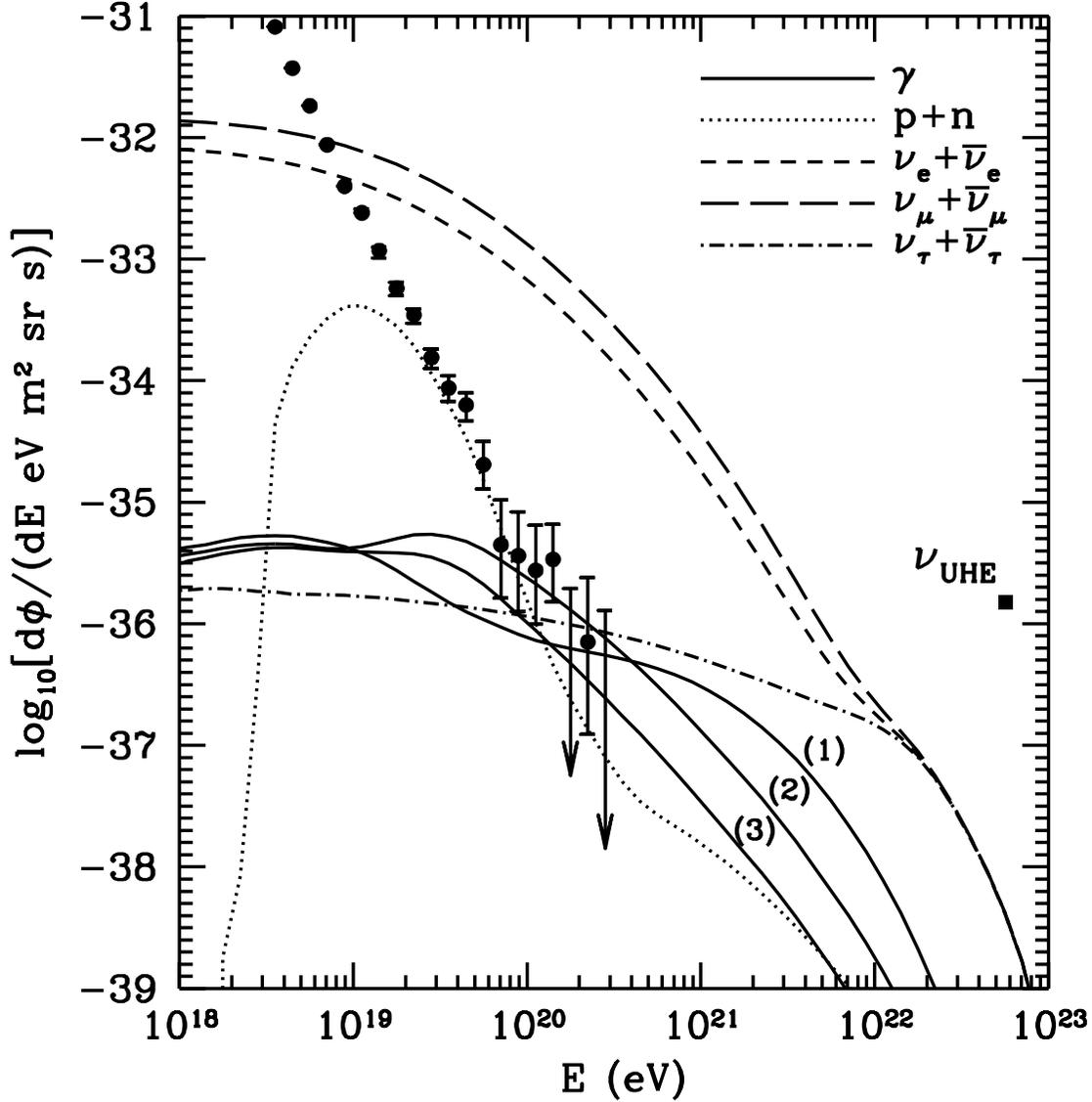}
\caption[]{Results of the simulation\cite{GK3} of ``$Z$ bursts" homogeneously 
distributed 
up to a redshift of 4. The points  and error bars are the AGASA data. The point 
labelled with $\nu_{UHE}$
 shows the needed flux of ultrahigh energy neutrinos at 
the Z-resonance energy, with no relic lepton asymmetry  
(it could be reduced by at most $1/20$ with large lepton asymmetries).}
%\label{eps1.1}
\end{figure}
shows the
result of this simulation with an approximate fit to the AGASA cosmic ray data.
The photon lines labelled 1, 2 and 3 correspond to three different radio 
background models, which we take to well represent  the uncertainty in these 
models. 
 The fit to the AGASA data provides the normalization of the photon and nucleon 
fluxes that allows us to determine the (assumed homogeneous
and isotropic) flux of ultra-high energy neutrinos at the $Z$-resonance energy
($0.6 \times 10^{23}$ eV) to be 
\begin{equation}
F_{\nu_{\rm UHE}} = 1.5~10^{-36} \frac{1}{eV m^2 sr sec} ,
\label{Fnu}
\end{equation}
if no lepton asymmetry is assumed.  With a large lepton asymmetry
this flux could be reduced by up to a factor of about 1/20.  

Without making an assumption on the spectrum of $\nu_{\rm UHE}$, the ``$Z$-burst" 
mechanism determines the $\nu_{\rm UHE}$ flux only at one point, the 
$Z$-resonance energy $E_{Res}$.  It is interesting to point out that just on the 
basis of that one point, at $0.6 \times 10^{23}$ eV, taking  the results 
of the ``Goldstone Experiment"\cite{GLN} at face value, the model of 
``$Z$-bursts" 
with relic  neutrinos of mass $m_{\rm SK}$ would  be already rejected, except for 
large lepton asymmetries.

The ``Goldstone Experiment" searches for Lunar radio emissions from
interactions of neutrinos (and cosmic rays) above $10^{19}$ eV of energy.
The published preliminary results\cite{GLN} correspond to only 12 hours of 
observation and systematic errors affecting the bounds are not yet well 
understood.  However, the present bounds show that the ``Goldstone Experiment"
will provide important constraints on ``$Z$-burst" models in the near future.

Finally, let us address the issue of how the huge lepton asymmetries mentioned
here could arise in the early Universe. 
Let us recall that, while charge neutrality imposes a lepton number
asymmetry in electrons as large as the baryon asymmetry in protons, i.e.
$(n_e-n_{\bar e}/n_\gamma)= (n_B-n_{\bar B}/n_\gamma) \simeq 10^{-10}$, no such 
restrictive bound operates on neutrinos.
 
 A realistic model to produce very large lepton asymmetries without producing a 
large baryon asymmetry was presented in Ref.\cite{CCG}.  The model requires the 
existence of bosons carrying lepton number (as in supersymmetric models), a 
period of inflation ending at a relatively low temperature, and a lepton 
asymmetry large enough  for the electroweak symmetry to be spontaneously
broken at all temperatures after inflation, which suppresses sphaleron 
transitions. Sphaleron transitions violate baryon plus lepton number ($B+L$)
while preserving ($B-L$), with the effect of producing $B=L$ if they are in 
equilibrium. Thus, in the presence of a very large lepton number $L$,  sphaleron 
transitions in equilibrium would produce an equally large baryon number $B$ and 
this needs to be avoided in these models.

 A model along similar lines is in Ref.\cite{D}.  Lepton asymmetries can also be 
generated in neutrino oscillations after the electroweak phase 
transition\cite{BD}, but it is unclear if an asymmetry of
order one can arise in this fashion\cite{Foot}, \cite{Dolgov}.

{\bf Acknowledgements-} This work was supported in part by the U.S. Department of 
Energy Grant No. DE-FG03-91ER40662, Task C.


\begin{thebibliography}{99}

\bibitem{SK} (Super-Kamiokande coll.) K. Scholberg, hep-ph/9905016.

\bibitem{Gawiser} E. Gawiser and J. Silk, {\it Science} {\bf 280}, 1405
(1988), astro-ph/9806197.


\bibitem{GK1} G. Gelmini and A. Kusenko, Phys. Rev. Lett. {\bf 82}, 5202
(1999); hep-ph/9902354.

\bibitem{AS} J. Adams and S. Sarkar, talk presented at the ``ICTP Workshop on
Physics of Relic Neutrinos", Trieste, Italy, September 1998.

\bibitem{PK} P. Pal and K. Kar, Phys. Lett. {\bf B451} (1999) 136, 
hep-ph/9809410.

\bibitem{FKT} K. Freese, E. W. Kolb, and M. S. Turner, Phys. Rev. {\bf D27}
(1983) 1689.

\bibitem{KS} H. Kang and G. Steigman, Nucl. Phys. B{\bf 372} (1992) 494.

\bibitem{AF} K. Abazajian and G. Fuller, private communication.

\bibitem{LP1} J. Lesgourgues and S. Pastor, Phys. Rev. {\bf D60}, 103521 (1999),
astro-ph/9904411.

\bibitem{KR} W. H. Kinney and A. Riotto Phys. Rev. Lett. {\bf 83} (1999) 17, 
hep-ph/9903459.

\bibitem{Wagoner} R.V. Wagoner, W.A. Fowler and F. Hoyle, Astro. Phys. J {\bf 
148},3 (1967).

\bibitem{W} T. Weiler, Astropart. Phys. {\bf 11}, 303 (1999).  See also
D. Fargion, B. Mele and A. Salis, Astrophys. J. {\bf 517}, 725 (1999).

\bibitem{bounds} E. Waxman and J. Bahcall, Phys. Rev. {\bf D59}, 023002
(1999); J. J. Blanco-Pillado, R. A. Vasquez and E. Zas, astro-ph/9902266.

\bibitem{GK2} G. Gelmini and A. Kusenko, Phys. Rev. Lett. {\bf 84}, 1378 (2000), 
 hep-ph/9908276.
 
\bibitem{sigl} G.~Sigl, S.~Lee, P.~Bhattacharjee and S.~Yoshida,
 Phys. Rev.  {\bf D59}, 043504 (1999).

\bibitem{CDF}J. Crooks, J. Dunn and P. Frampton, astro-ph/0002089.

\bibitem{S1} G. Sigl (private communication).


\bibitem{GK3} G. Gelmini, A. Kusenko, S. Nussinov, and G. Varieschi, in
preparation.

\bibitem{BS} P. Bhattacharjee and G. Sigl, Phys. Rept. {\bf 327}, 109 (2000).

\bibitem{PB} R. J. Protheroe and P. L. Biermann, Astropart. Phys. {\bf 6}, 45
(1996), Erratum-{\it ibid} {\bf 7}, 181 (1997).

\bibitem{GLN} P. W. Gorham, K. M. Liewer and C. J. Naudet, astro-ph/9906504.

\bibitem{CCG} A. Casas, W. Y. Cheng and G. Gelmini, Nucl. Phys. {\bf B538}, 297
(1999), hep-ph/9709289.

\bibitem{D} J. McDonald, hep-ph/9908300.

\bibitem{BD} R. Barbieri and A. Dolgov, Nucl. Phys. {\bf B237}, 742 (1991).

\bibitem{Foot} R. Foot, M. Thompson and R. R. Volkas, Phys. Rev. {\bf D53}, 5349 
(1996); X. Shi, Phys. Rev. {\bf D54}, 2753 (1996), A. K. Dolgov {\it et al.}
hep-ph/9910444.

\bibitem{Dolgov} A.D. Dolgov, S.H. Hansen, S. Pastor, D.V.
Semikoz, hep-ph/9910444.



\end{thebibliography}
\end{document}